\documentclass[11pt,twoside]{article}
\usepackage{asp2004}
\usepackage{psfig}
\usepackage{epsf}
\usepackage{graphics}
\usepackage{lscape}
\def\emphasize#1{{\itshape#1\/}}
\def\arg#1{{\it#1\/}}
\let\prog=\arg
\def\edcomment#1{\iffalse\marginpar{\raggedright\sl#1\/}\else\relax\fi}
\reversemarginpar

\begin{document}
\title{Feedback and Formation of Massive Star Clusters in Giant Molecular Clouds}
\author{Jonathan C. Tan$^1$ and Christopher F. McKee$^2$}
\affil{$^1$ Princeton University Observatory, Princeton, NJ 08544, USA\\ $^2$ Depts. of Physics and Astronomy, UC, Berkeley, CA 94720, USA}

\begin{abstract}
  It is well known that the energy input from massive stars dominates
  the thermal and mechanical heating of typical regions in the
  interstellar medium of galaxies. These effects are amplified
  tremendously in the immediate environment of young massive star
  clusters, which may contain thousands of O and B stars. We present
  models of star cluster formation that attempt to account for the
  interplay between feedback and self-gravity in forming clusters,
  with application to systems ranging from the Orion Nebula Cluster,
  the Galactic Center clusters like the Arches, and super star
  clusters.
\end{abstract}
\thispagestyle{plain}

\section{The Initial Conditions for Galactic Star Cluster Formation}

Star formation in disk galaxies appears to be dominated by a clustered
mode (Lada \& Lada 2003), occurring in regions of the disk that are
gravitationally unstable (Martin \& Kennicutt 2001). The most unstable
mass scales are about $1.3\times 10^7 (\sigma_{\rm gas}/6\;{\rm
  km\:s^{-1}})^4 (\Sigma_{\rm gas}/10\;M_\odot {\rm pc}^{-2})^{-1}
M_\odot$, roughly consistent with the observed cutoff of the Galactic
Giant Molecular Cloud (GMC) mass function (Williams \& McKee 1997).
%cfm
GMCs are gravitationally bound (Solomon et al. 1987), and
there is a comparable amount of atomic gas associated with these structures
(Blitz 1990). Altogether, a
%cf
significant fraction ($\sim 1/2$) of the total gas mass inside the
solar circle is organized into bound structures.

The lifetimes of GMCs are not well known. 
%cfm
Blitz \& Shu (1980) used observational
and theoretical arguments to infer a GMC lifetime of
a few times $10^7$~yr; their argument based on 
photoevaporation of the clouds was confirmed in
a more extensive calculation by Williams \& McKee (1997).
The rocket effect associated with the photoevaporation
displaces the molecular gas from regions of active star
formation at the same time the gas is being transformed
from molecular form to atomic and ionized form.
%cf
Leisawitz, Bash \& Thaddeus
(1989) found that open star clusters older than about $\sim 10$~Myr
were not associated with molecular clouds, which is consistent either
with post-star-formation cloud lifetimes shorter than this age or with
relative velocities of star clusters and their parent clouds of about
$10\:{\rm km\:s^{-1}}$.  
Elmegreen (2000) presented a number of arguments for star formation to
occur in 
%cfm
about $1-2$
dynamical timescales, 
%cf
particularly from the
small age spreads in clusters such as the Orion Nebula Cluster and
from the statistics of the presence of young stars in regions of dense
gas. 
%cfm
Hartmann (2003) has argued for such rapid star formation
in Taurus.
%cfm I suggest that we omit the damping arguments, since
%       as you point out we need to have additional sentences, and we
%       do not have space.
%Numerical models of MHD turbulence in GMCs find
%decay timescales of only $\sim$ a dynamical time (Stone, Ostriker \&
%Gammie 1998; MacLow et al. 1998). Stone et al. (1998) estimate that
%without additional driving, turbulence will decay in GMCs in $\sim
%5-10$~Myr.
%???Vazquez-Semadeni et al. (2000) used
%this result to argue that clouds must be younger than a flow crossing
%time, and so again have lifetimes of only a few~Myr.
%However, it is not clear than the decay result is secure (Lazarian ???). 
%cfm Following is irrelevant, so skip it:
%Given the approximately equal mass fractions of bound and unbound gas,
%the formation timescale of bound clouds should be about equal to their
%destruction times, i.e. an individual atom spends roughly equal times
%in bound and unbound states. 
On the other hand, the observation that
the angular momentum vectors of GMCs
in M33 are small and are both pro- and retro-grade (Rosolowsky et al. 
2003) may indicate a relatively long
lifetime so that angular momentum can be shed by magnetic braking
(e.g., Mestel 1985)
and/or cloud-cloud collisions in a shearing disk (Tan 2000).
%cf

%cfm
Most star formation in GMCs
is concentrated in {\it clumps} that occupy a relatively small
fraction of the volume and have a small fraction of the mass.
%cfm a protocluster is stellar, whereas a clump is gaseous
%or {\it protoclusters} that eventually become star clusters.  
%cf
Within
clumps are over-dense regions that we refer to as {\it cores} that
form individual stars or binaries
(Williams, Blitz, \& McKee 2000). Star-forming clumps are identified
by $\rm H_2O$ masers, outflows, and far infrared and radio continuum
emission (Plume et al.  1997; Hunter et al. 2000; Sridharan et al.
2002; Beuther et al. 2002; Zhang et al. 2002; Mueller et al. 2002;
Shirley et al. 2003). Most of the 350~$\rm \mu m$ emission maps of
Shirley et al. (2003) have morphologies consistent with
quasi-spherical, virialized distributions.  Typical properties of
clumps are masses $M\sim 100-10^4\: M_\odot$, diameters $\sim 1$~pc
and surface densities $\Sigma\sim 1\:{\rm g\:cm^{-2}}$.

Clump properties are very similar to those of 
%cfm
the smaller
%cf
infrared dark clouds
(IRDCs) (Carey et al. 1998), which are dense, cold regions embedded in
GMCs. Thus, understanding the origin of IRDCs is crucial for progress in
the fields of both star cluster formation and galaxy formation and
evolution. Carey et al. noted nongaussian emission line profiles of
$\rm H_2CO$ from 9 out of 10 IRDCs. This may indicate short formation
timescales via a triggering mechanism. IRDC morphologies are also more
filamentary than the sub-mm emission maps of star-forming clumps,
which suggest that the latter represent a more evolved
stage.

Possible triggers for IRDC formation include compression of parts of
GMCs (which are likely to contain clumpy substructure) by shock waves
driven by cloud-cloud collisions (Scoville, Sanders \& Clemens 1986;
Tan 2000), spiral density waves (Sleath \& Alexander 1996), ionization
fronts (Elmegreen \& Lada 1977; Thompson et al. 2004), supernovae
(e.g. Palous, Tenorio-Tagle \& Franco 1994) or stellar winds (e.g.
Whitworth \& Francis 2002). Clumps that were previously
pressure-confined and gravitationally stable may suddenly become
unstable.
%cfm But if B does not dominate gravity initially, it never will
%Compression raises magnetic pressure
%support, which has contributions from a background field and a
%turbulent field (Zweibel \& McKee 1995; McKee \& Tan 2003).  
The contraction of the cloud can be halted by the onset
of star formation (McKee 1989).  
%cf
We expect that the
protocluster will come into an approximate equilibrium in which
pressure support balances self-gravity.
Some pieces of evidence in support of this view are 
the approximately spherical
morphologies of star-forming clumps (Shirley et al. 2003); the
timescales of star cluster formation estimated from outflow momentum
generation rates (Tan \& McKee 2002); and the empirical age spreads of
stars in young clusters, such as the Orion Nebula Cluster (Palla \&
Stahler 1999), which are quite long compared to the dynamical or
free-fall timescales of clumps, $\bar{t}_{\rm ff}=(3\pi/32G\bar{\rho})^{1/2}
=1.0\times10^5
(M/4000M_\odot)^{1/4}\Sigma^{-3/4}\:{\rm yr}$.
%cfm CHECK THE FOLLOWING CAREFULLY. GIVE THE VALUE OF t_sf FOR THE ONC
%       FROM PALLA AND STAHLER
This last point requires comment, since Elmegreen (2000) uses
the Orion Nebula Cluster as an example of star formation
on a dynamical time. He estimated the density in the cluster prior to
star formation as $n_{\rm H}=1.2\times 10^5$ cm$^{-3}$, corresponding
to a free-fall time of $1.25\times 10^5$ yr. If the star formation
occurred over a time $t_{\rm sf}=10^6$ yr, 
then $\eta\equiv t_{\rm sf}/\bar t_{\rm ff}
\simeq 8\gg 1$. He used the dynamical time $t_{\rm dyn}\equiv
R/\sigma$ to compare with $t_{\rm sf}$.  If the virial
parameter $\alpha_{\rm vir}\equiv 5\sigma^2 R/GM\sim 1$,
as observed in star-forming regions (McKee \& Tan 2003),
then $R/\sigma =2.0\bar t_{\rm ff}/\alpha_{\rm vir}^{1/2}
\simeq 2.5\times 10^5$ yr. For $t_{\rm sf}\simeq 10^6$~yr,
star formation in the Orion Nebula Cluster was rapid, but
not so rapid that a quasi-equilibrium treatment is invalid.
%cf

\section{Pressurized Star and Star Cluster Formation}

McKee \& Tan (2002; 2003) considered star formation from marginally
unstable, turbulent gas cores that are embedded in high
pressure regions, such as the centers of star-forming clumps where
$P\simeq 1.8G\Sigma^2 = 8.5\times 10^9 \Sigma^2 \:{\rm K\:cm^{-3}}$.
High pressures cause equilibrium cores to be compact, $r_{\rm core}
\simeq 0.06 (m_{*f}/30M_\odot)^{1/2}\Sigma^{-1/2}\:{\rm pc} $, and
collapse times short, $t_{*f}\simeq 1.3\times 10^5
(m_{*f}/30M_\odot)^{1/4} \Sigma^{-3/4}\:{\rm yr}$, where $m_{*f}$ is
the final stellar mass forming from the core with 50\% efficiency.
These properties help to overcome some objections to core-based models
of massive star formation (Stahler, Palla \& Ho 2000; Tan 2003). The
model helps to understand features of the Orion hot core
(Tan 2004), a close example of a massive protostar.

We have applied the above {\it turbulent core} star formation model to
star clusters (Tan \& McKee 2002). As a first approximation we assume
that stars form independently from one another. This ignores the
effects of mutual stellar interactions (e.g. Ostriker 1994; Elmegreen
\& Shadmehri 2003; Bonnell, Bate \& Vine 2003). 
%cfm Suggest omitting to save space:
%For present purposes
%we regard these as being relatively small perturbations to a basic
%model of formation from distinct cores, although the issue deserves
%further study. 
%cf
The simplicity of our approach means we only need to
specify a stellar initial mass function (IMF) and an overall star
formation rate to define a particular star cluster formation model. We
use an empirical IMF in the form of a Salpeter power law 
(${\rm d}{\cal N}/{\rm d\:ln}\; m\propto m^{-1.35}$) at the high-mass end.
% and allow for a break around a solar mass. 
%cfm
We parameterize the star-formation rate 
in terms of $\eta\equiv t_{\rm sf}/\bar t_{\rm ff}$, the number
of free-fall
timescales it takes to form the stars. We adopt
a star-formation efficiency of 0.5. 
%cf
Stars are drawn at random from the IMF according to this
rate and their individual protostellar evolution is followed. The
properties of all the individual stars, such as bolometric and
ionizing luminosities and protostellar outflow momentum flux, are
summed up to give the overall cluster properties.  The effects of
discreteness of the IMF are easily probed via Monte Carlo simulations
of many star clusters.

These models were used to estimate the expected protostellar outflow
momentum flux from forming clusters as a function of the star
formation rate. Most models of outflows (e.g. K\"onigl \& Pudritz
2000; Shu et al. 2000) predict a linear relation between the accretion
and outflow rates from protostars. The outflow velocity is of order
the escape speed from the stellar surface.  From a comparison to a
sample of observed cluster outflows we estimated that $\eta\sim 10$,
but with large uncertainties (Tan \& McKee 2002).

\section{Feedback Processes in Star Cluster Formation}

\begin{figure}[!ht]
%clusterfeedback
\plotone{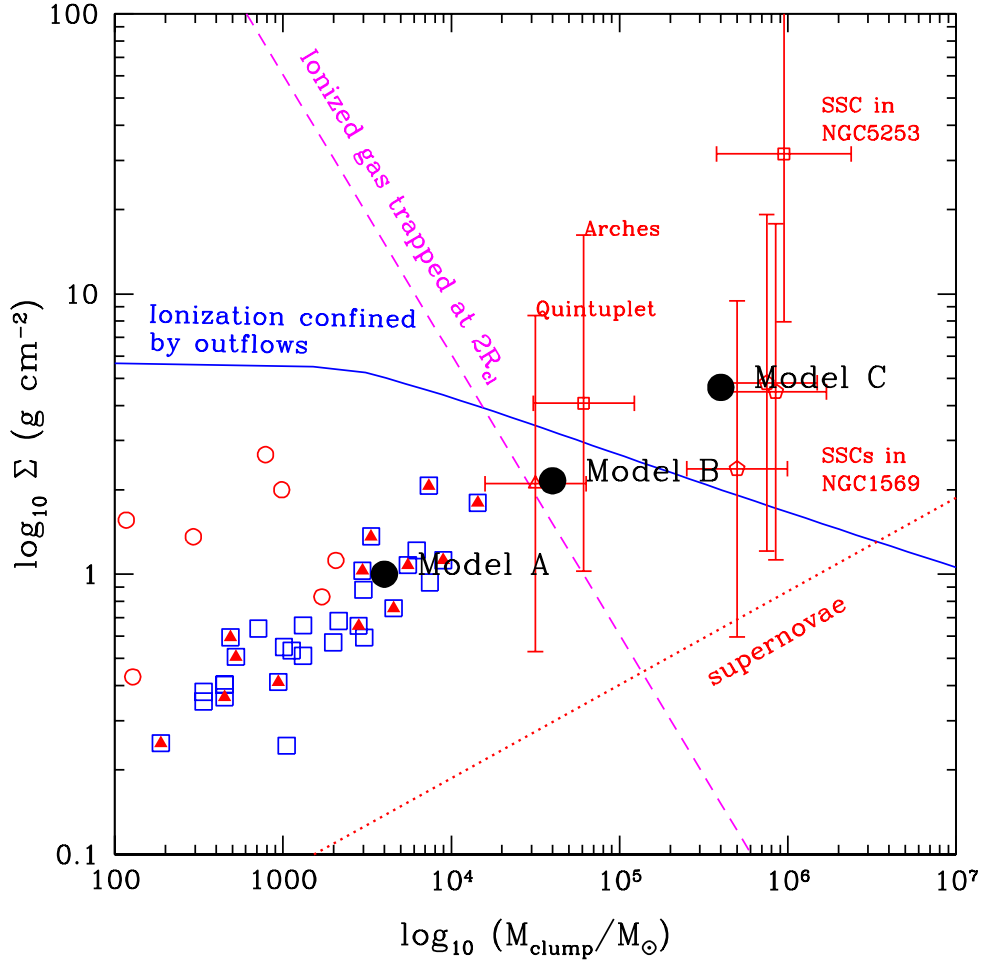}
%\plottwo{}{}
%\plotfiddle{file}{vsize}{rot}{hsf}{vsf}{htrans}{vtrans}
\caption{
  Surface Density vs. Mass diagram for star cluster formation.  To the
  right of the dashed line the escape velocity from twice the clump
  radius is greater than the ionized gas sound speed ($\sim 10\:{\rm
    km\:s^{-1}}$)
%($11.4\:{\rm km\:s^{-1}}$ for gas at $T=10^4\:{\rm K}$) 
  and ionizing feedback is less effective. The dotted line (of
  constant volume density) shows when the formation time, $\eta
  \bar{t}_{\rm ff}$, equals 3~Myr, assuming $\eta=10$. Below this line
  supernova feedback becomes important.  Above the solid line,
  ionizing photons from a typical sampling of the stellar IMF are
  confined by a protocluster wind formed from the combination of
  individual protostellar outflows (Tan 2001), again assuming
  $\eta=10$. The open circles are the outflow sample used in Tan \&
  McKee (2002), the open squares are the sample of Mueller et al.
  (2002), with solid triangles indicating presence of an HII region.
  The Arches and Quintuplet Galactic center clusters and several SSCs
  are shown, including corrections for formation efficiency. The
  solid circles are theoretical feedback models (Tan \&
  McKee 2001 and shown in Fig.~3b).}
\end{figure}

\begin{figure}[!ht]
%\plotone{modelA.ps}
%\plottwo{}{}
%modelC
\plotfiddle{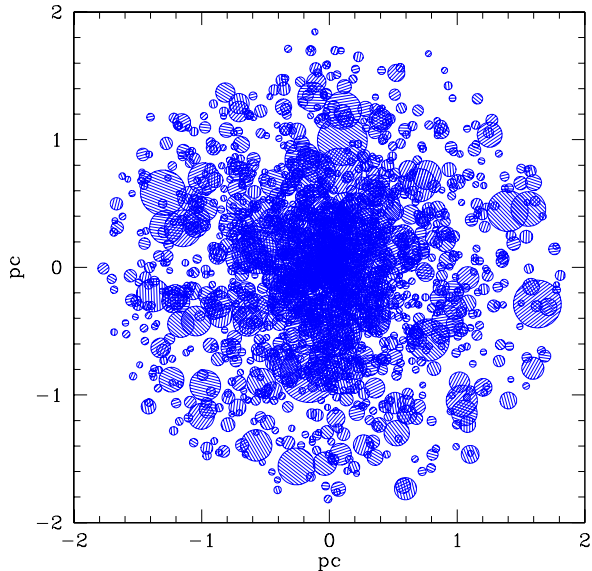}{2.5in}{0}{40}{40}{-125}{-75}
%{}{rot}{hsf}{vsf}{htrans}{vtrans}
\caption{Projected initial density structure of proto-SSC medium (model~C), consisting of cores embedded in an intercore medium (not shown).}
\end{figure}

\begin{figure}[!ht]
%cldesttheory and cldestnew
\plottwo{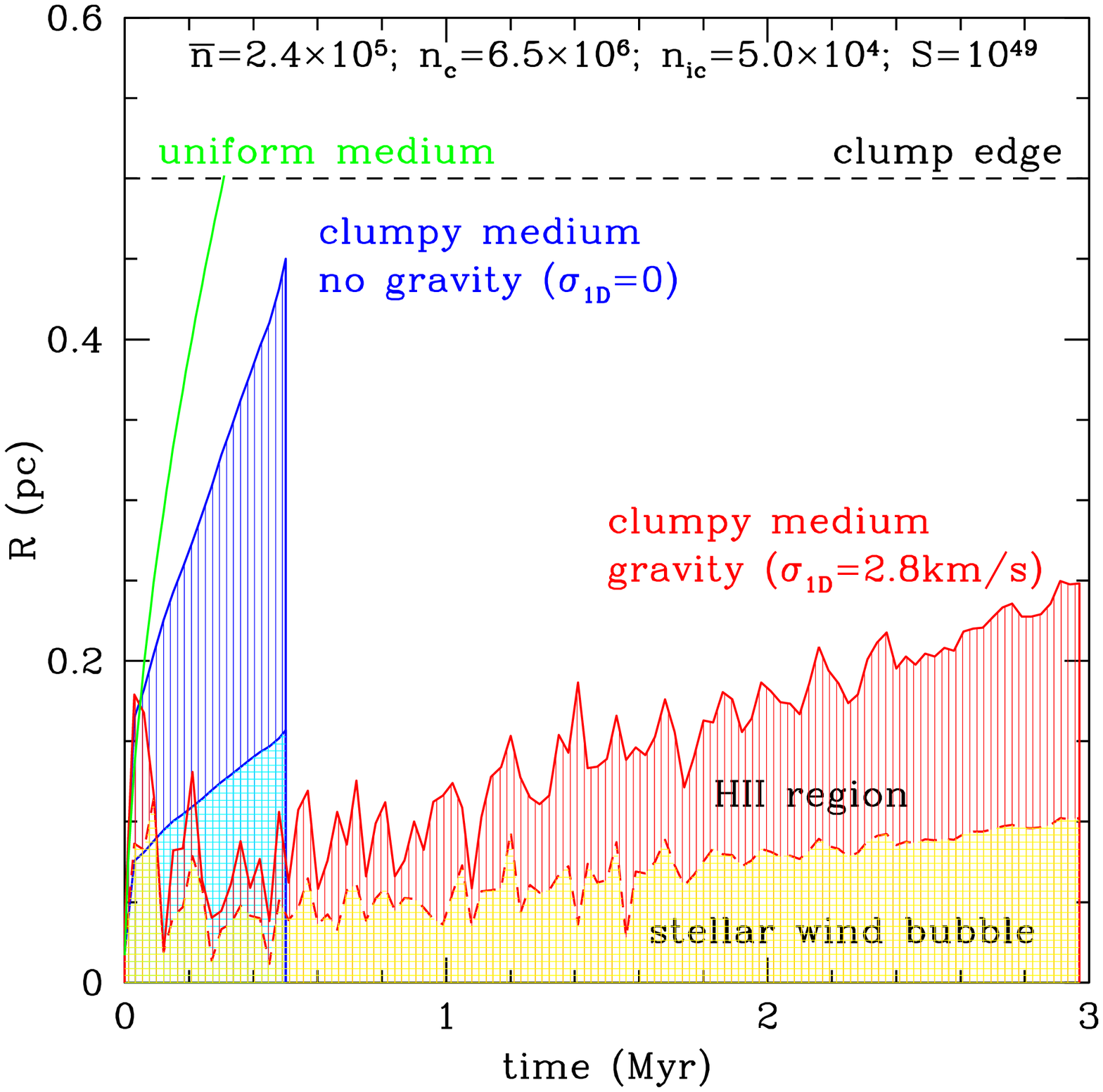}{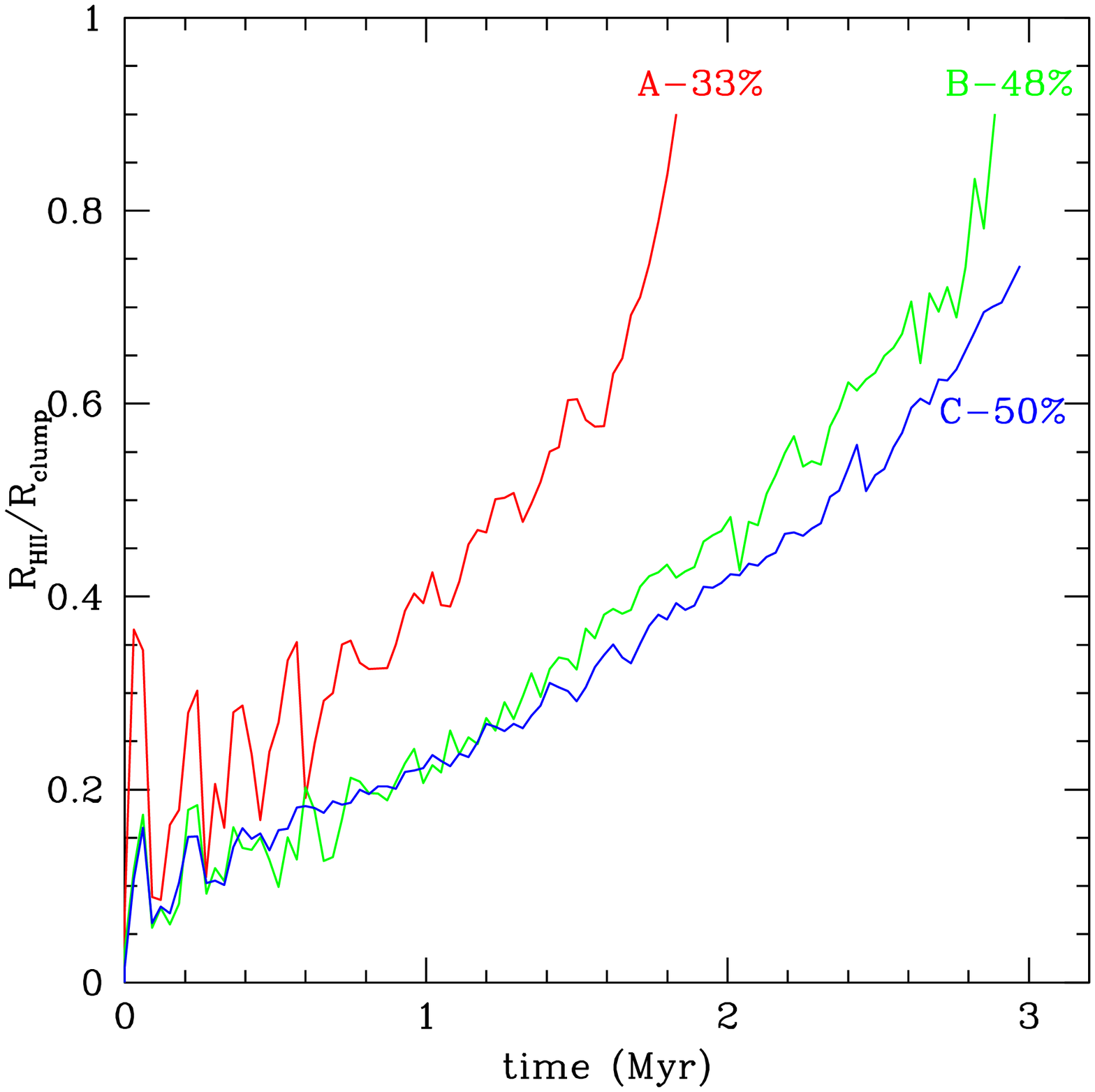}
\caption{ 
  Evolution of feedback in protoclusters. (a) Left panel: test case
  showing size evolution of the HII region and stellar wind bubble
  created by a single O~star with ionizing photon luminosity
  $S=10^{49}\:{\rm s}^{-1}$ embedded in a typical Galactic
  protocluster (model A), with mean, core and intercore densities
  indicated in units of $\rm H\:cm^{-3}$. In a uniform medium at the
  mean density the HII region expands to the clump edge in just over
  $10^5$~yr. If the medium is clumpy the feedback progresses more
  slowly because ionization fronts tend to encounter much denser gas
  with short recombination times and inner cores shield outer regions.
  In this modeling we treat the HII region and wind bubble in one
  dimension, averaging over angular variations.  The thermal pressure
  of the HII region confines the ram pressure of the stellar wind
  bubble, which is shown by the inner shaded region.  Break-out occurs
  after about $6\times 10^5$~yr.  Finally, including self-gravity so
  that the cores have turbulent motions and the protocluster is in
  virial equilibrium, feedback is severely confined for at least
  3~Myr. It is harder to push the cores out of the potential of the
  clump and their motions lead to the continuous replenishment of
  neutral gas in the HII region. (b) Right panel: evolution of HII
  region size for clumpy, self-gravitating models A, B and C for
  protoclusters, forming with $\eta=30$. The final star formation
  efficiency of each model is shown, assuming formation stops once the
  HII region reaches the edge of the clump or after 3~Myr by the
  action of supernovae.  Non-supernova feedback is quite weak in
  models B and C and the embedded phase lasts several Myr.}
\end{figure}

Feedback processes that act against gravitational collapse and
accretion of gas to protostars include radiation pressure (transmitted
primarily via dust grains), thermal pressure of ionized regions and
ram pressure from stellar winds, particularly MHD-driven outflows from
protostars that are still actively accreting. 
%cfm This is complicated, and since it is not necessary for the
%       argument I have dropped it.
%The pressure support of
%the protocluster before the onset of star formation is likely to be
%dominated by magnetic pressure, including contributions from both
%large-scale and small-scale (turbulent) fields that couple to the gas
%by the small fraction of ions. A clump that is about to form a star
%cluster must be overall magnetically super-critical so that magnetic
%stresses are overwhelmed by gravity. However, it is possible that once
%stars are forming the balance changes as both new flux is generated
%(e.g. from MHD outflows) 
%cfm Note that coupling is irrelevant to pressure
%and the coupling is enhanced via greater
%levels of ionization (e.g. from X-rays). Similarly, 
Pressure support
from turbulent bulk motions of the gas and the turbulent
magnetic field is likely to be quite important
in the initial clump (McKee \& Tan 2003). Indeed 
the decay of this turbulence may be one
of the most important factors controlling the onset of star cluster
formation (McKee 1989). 
%cf
Forming stars in the protocluster can rejuvenate the
turbulence, primarily by the momentum input from their MHD outflows.

Tidal torques among cores and their formation from a turbulent medium
lead to a distribution of initial angular momenta.  This
distribution and the effectiveness of transport processes may also be
important in regulating the cluster star formation rate.  Important
transport processes include magnetic braking (Mestel 1985; Allen, Li
\& Shu 2003), the magneto-rotational (Balbus \& Hawley 1991; Salmeron
\& Wardle 2003) and gravitational instabilities (Larson 1984; Johnson
\& Gammie 2003) in disks, including global modes such as spiral
density waves (Lynden-Bell \& Kalnajs 1972) that may be easily
induced by the presence of binaries or massive planets (Blondin 2000).
We shall assume that these processes are efficient enough not to
impede stellar accretion.

The complexity of the feedback processes and their interplay with a
turbulent, self-gravitating medium forces us to make several
simplifying assumptions in our modeling. Our goal is to explore which
processes and aspects of the problem are most important.  First we
layout the parameter space of cluster mass and surface density
(Fig.~1). As described in the caption, we can make simple estimates of
when various feedback effects are important: e.g. the dashed line
indicates when the escape speed from twice the protocluster radius is
the ionized gas sound speed. Using the adopted empirical value for the
overall star formation rate of $\eta=10$, we show the condition
for star formation to be complete within 3~Myr (i.e. before supernova
feedback is important). With this rate we can also estimate the
combined properties of a protocluster wind composed of superposed MHD
outflows. This wind can be dense enough to confine ionizing feedback.
Various observed star clusters are also shown in Fig.~1.

To move beyond these simple estimates we have constructed a dynamical
feedback model (Tan \& McKee 2001). The key element here is to
investigate the effect of a turbulent and clumpy medium. We
approximate this structure by dividing the gas into a population of
cores with 80\% of the total mass and an intercore medium. The cores
have a mass spectrum ${\rm d}{\cal N}/{\rm d\:ln}\; m\propto m^{-0.6}$, a
uniform density, and a centrally concentrated initial spatial
distribution (Fig.~2). The dynamics of the cores are affected by the
potential of the overall protocluster and feedback effects from a
stellar population at the cluster center. These include radiation
pressure, stellar winds and ionization, which can photoevaporate cores
according to the models of Bertoldi (1989) and Bertoldi \& McKee
(1990). 
%cfm We already said this:
%For more details see Tan \& McKee (2001).
%cf

Figure~3a shows the results of a test case for a Galactic protocluster
($4000M_\odot$, $\Sigma=1\:{\rm g\:cm^{-2}}$, model A in Fig.~1)
interacting with the feedback from a single O star. As described in
the caption, a clumpy, turbulent medium is much more capable of
confining feedback. Such effects are undoubtedly important for
confinement of ultra-compact HII regions, allowing them to be
relatively long-lived.

Figure~3b shows the evolution of the size of the HII region for the
three cases A, B and C shown in Fig.~1, which are relevant to typical
Galactic massive clusters, Galactic Center clusters like the Arches,
and super star clusters (SSCs). Note these models correspond to a
constant volume density. A relatively slow star formation rate of
$\eta=30$ was adopted and material for new stars was taken from the
innermost cores. Starburst99 models (Leitherer et al. 1999) were used
for the mean stellar properties.  Again feedback effects are weaker
than commonly supposed. In cases B and C pre-supernova feedback is
confined for $\sim$3~Myr so the embedded phase lasts for at least this
long. We predict that Wolf-Rayet spectral features should be seen in
%cfm ADDED YOUNG AND WAFFLED:
most young, 
%cf
optically visible SSCs. There are also implications for the
numbers of observed embedded clusters (as probed in radio continuum,
e.g.  Johnson \& Kobulnicky 2003) relative to young optically-revealed
systems.

Even in the Galactic case (model A), which has an escape speed small
compared to the ionized gas sound speed, ionization is confined for
almost 2~Myr.  This may be relevant to the Orion Nebula Cluster where
stars still form in close proximity to O stars and where some
cluster members are thought to have ages of $\sim 2$~Myr (Palla \&
Stahler 1999; Hoogerwerf, de Bruijne, \& de Zeeuw 2001).

We emphasize that several important physical processes are not
included in these models. Our treatment of stellar winds does not
include protostellar MHD outflows or Wolf-Rayet winds, so wind
feedback is underestimated. We assume a central point-like stellar
distribution that enhances wind feedback because there are no
wind-wind shocks. We ignore dynamical ejection of massive stars, which
seems to be a relatively efficient: Hoogerwerf et al. (2001)
argue that 4 out of 10 stars with $m_*>10M_\odot$ have been
ejected from the Orion Nebula Cluster. This obviously reduces the
effectiveness of feedback. We assume star formation only begins once
the clump is at the chosen density: in reality there will be a more
gradual onset.  We ignore further accretion to the clump from
the surrounding GMC. The two component description of the
gas is highly approximate.

The complexity of the star cluster formation process is daunting. 
%cfm revised
The models we have developed are clearly a very crude first step.
%cf
The reliability of such models
can be established only with very detailed comparisons to observations
of real clusters such as the Orion Nebula Cluster, R136 in 30~Doradus
and nearby SSCs. For such comparisons, improved methods to date stars that are younger than
a few Myr are needed.

\acknowledgements We are supported by a Princeton Spitzer-Cotsen fellowship
\& NASA grant NAG5-10811 (JCT) \& NSF grant
AST-0098365 (CFM).
% and by a NASA grant supporting the Center for Star
%Formation Studies.

\end{document}